\documentclass[sigconf,nonacm]{acmart}

\usepackage{graphicx}
\usepackage{color,soul}
\usepackage{xspace}
\usepackage[utf8]{inputenc}
\usepackage{listings}
\usepackage{xurl}
\usepackage{hyperref}
\usepackage{fmtcount}
\usepackage[most]{tcolorbox}
\usepackage[frozencache]{minted2}
\usepackage{changepage}

\newenvironment{quotationsm}{%
	\par\addvspace{\topsep}%
	\begin{adjustwidth}{0.25cm}{0.25cm}%
	\begin{quote}%
}{%
	\end{quote}%
	\end{adjustwidth}%
	\par\addvspace{\topsep}%
}

\newcommand{\teama} {{\scshape T01}\xspace}
\newcommand{\teamb} {{\scshape T02}\xspace}

\newcommand{\teamd} {{\scshape T04}\xspace}

\newcommand{\teamg} {{\scshape T07}\xspace}

\newcommand{\teamj} {{\scshape T10}\xspace}
\newcommand{\teamk} {{\scshape T11}\xspace}
\newcommand{\teaml} {{\scshape T12}\xspace}
\newcommand{\teamm} {{\scshape T13}\xspace}
\newcommand{\teamn} {{\scshape T14}\xspace}
\newcommand{\teamo} {{\scshape T15}\xspace}

\newcommand{\teamr} {{\scshape T18}\xspace}

\newcommand{\teamt} {{\scshape T20}\xspace}
\newcommand{\teamu} {{\scshape T21}\xspace}

\newcommand{\parta} {{\scshape C01-T08}\xspace}
\newcommand{\partb} {{\scshape C02-T10}\xspace}
\newcommand{\partc} {{\scshape A03-T11}\xspace}
\newcommand{\partd} {{\scshape C04-T04}\xspace}
\newcommand{\parte} {{\scshape B05-T18}\xspace}
\newcommand{\partf} {{\scshape B06-T22}\xspace}
\newcommand{\partg} {{\scshape C07-T10}\xspace}
\newcommand{\parth} {{\scshape B08-T07}\xspace}
\newcommand{\parti} {{\scshape A09-T10}\xspace}
\newcommand{\partj} {{\scshape B10-T07}\xspace}
\newcommand{\partk} {{\scshape B11-T12}\xspace}
\newcommand{\partl} {{\scshape B12-T23}\xspace}
\newcommand{\partm} {{\scshape C13-T24}\xspace}
\newcommand{\partn} {{\scshape A14-T06}\xspace}
\newcommand{\parto} {{\scshape B15-T25}\xspace}
\newcommand{\partp} {{\scshape A16-T09}\xspace}
\newcommand{\partq} {{\scshape B17-T13}\xspace}
\newcommand{\partr} {{\scshape C18-T08}\xspace}
\newcommand{\parts} {{\scshape B19-T15}\xspace}
\newcommand{\partt} {{\scshape B20-T12}\xspace}
\newcommand{\partu} {{\scshape B21-T13}\xspace}
\newcommand{\partv} {{\scshape B22-T26}\xspace}

\newcounter{recommendation}
\newcommand{\recommendation}[1]{%
  \stepcounter{recommendation}%
  \begin{tcolorbox}[%
	enhanced,
	breakable,
	colback={rgb,1:red,0.95;green,0.95;blue,1},
	boxsep=0pt,
	left=2.5pt,
	right=2.5pt,
	top=1pt,
	bottom=1.5pt,
	boxrule=1.0pt,
  ]
  \textbf{Recommendation \the\value{recommendation}:} #1
  \end{tcolorbox}%
}

\newcounter{comparison}
\newcommand{\comparison}[2]{
	\stepcounter{comparison}
	\begin{tcolorbox}[
		boxsep=0pt,
		left=2pt,
		right=2pt,
		top=1pt,
		bottom=1.5pt,
		boxrule=0.5pt,
	]
		\textbf{#1:}
		#2
	\end{tcolorbox}
}

\newcounter{observation}
\newcommand{\observation}[1]{
	\stepcounter{observation}
	\begin{tcolorbox}[
		enhanced,
		breakable,
		boxsep=0pt,
		left=2pt,
		right=2pt,
		top=1pt,
		bottom=1pt,
		boxrule=0.5pt,
	]
		\textbf{Observation \decimal{observation}:}
		#1
	\end{tcolorbox}
}

\AtBeginDocument{%
  }

\begin{document}

\title{%
  \texorpdfstring{%
	``We just did not have that on the embedded system'':\\
	Insights and Challenges for Securing Microcontroller Systems from the Embedded CTF Competitions%
  }{
	``We just did not have that on the embedded system'': Insights and Challenges for Securing Microcontroller Systems from the Embedded CTF Competitions%
  }%
}

\thanks{This is the authors' version of the paper accepted to appear in the Proceedings of the 2025 ACM SIGSAC Conference on Computer and Communications Security (CCS '25). The definitive version is available at https://doi.org/10.1145/3719027.3765039}

\author{Zheyuan Ma}
\orcid{0009-0009-5631-0466}
\affiliation{%
	\institution{CactiLab, University at Buffalo}
	\city{Buffalo}
	\state{NY}
	\country{USA}
}
\email{zheyuanm@buffalo.edu}
\author{Gaoxiang Liu}
\orcid{0009-0004-9304-8317}
\affiliation{%
	\institution{CactiLab, University at Buffalo}
	\city{Buffalo}
	\state{NY}
	\country{USA}
}
\email{gliu25@buffalo.edu}
\author{Alex Eastman}
\orcid{0009-0008-8774-8885}
\affiliation{%
	\institution{CactiLab, University at Buffalo}
	\city{Buffalo}
	\state{NY}
	\country{USA}
}
\email{alexeast@buffalo.edu}
\author{Kai Kaufman}
\orcid{0009-0002-4217-8897}
\affiliation{%
	\institution{Worcester Polytechnic Institute}
	\city{Worcester}
	\state{MA}
	\country{USA}
}
\email{ktkaufman@wpi.edu}
\author{Md Armanuzzaman}
\orcid{0009-0004-5264-7962}
\affiliation{%
	\institution{CactiLab, Northeastern University}
	\city{Boston}
	\state{MA}
	\country{USA}
}
\email{m.armanuzzaman@northeastern.edu}
\author{Xi Tan}
\orcid{0009-0001-6368-1523}
\affiliation{%
	\institution{SUNRISE Lab, University of Colorado Colorado Springs}
	\city{Colorado Springs}
	\state{CO}
	\country{USA}
}
\email{xtan4@uccs.edu}
\author{Katherine Jesse}
\orcid{0009-0001-8957-4507}
\affiliation{%
	\institution{Worcester Polytechnic Institute}
	\city{Worcester}
	\state{MA}
	\country{USA}
}
\email{kcjesse@wpi.edu}
\author{Robert J.\ Walls}
\orcid{0000-0002-1338-6403}
\affiliation{%
	\institution{Worcester Polytechnic Institute}
	\city{Worcester}
	\state{MA}
	\country{USA}
}
\email{rjwalls@wpi.edu}
\author{Ziming Zhao}
\orcid{0000-0002-4930-5556}
\affiliation{%
	\institution{CactiLab, Northeastern University}
	\city{Boston}
	\state{MA}
	\country{USA}
}
\email{z.zhao@northeastern.edu}

\renewcommand{\shortauthors}{Zheyuan Ma et al.}

\begin{abstract}
	Microcontroller systems are integral to our daily lives, powering mission-critical applications such as vehicles, medical devices, and industrial control systems.
	Therefore, it is essential to investigate and outline the challenges encountered in developing secure microcontroller systems.
	While previous research has focused solely on microcontroller firmware analysis to identify and characterize vulnerabilities, our study uniquely leverages data from the 2023 and 2024 MITRE eCTF team submissions and post-competition interviews. 
	This approach allows us to dissect the entire lifecycle of secure microcontroller system development from both technical and perceptual perspectives, providing deeper insights into how these vulnerabilities emerge in the first place.
	
	Through the lens of eCTF, we identify fundamental conceptual and practical challenges in securing microcontroller systems. 
	Conceptually, it is difficult to adapt from a microprocessor system to a microcontroller system, and participants are not wholly aware of the unique attacks against microcontrollers.
	Practically, security-enhancing tools, such as the memory-safe language Rust, lack adequate support on microcontrollers. 
	Additionally, poor-quality entropy sources weaken cryptography and secret generation.
	Our findings articulate specific research, developmental, and educational deficiencies, leading to targeted recommendations for researchers, developers, vendors, and educators to enhance the security of microcontroller systems.
\end{abstract}


\maketitle

\section{Introduction}

A microcontroller (MCU) is a compact, integrated circuit designed specifically for control tasks within embedded systems and Internet of Things (IoT) devices. 
Unlike general-purpose processors found in computers, a microcontroller integrates a processor core (CPU), memory (RAM and ROM), and peripheral interfaces---such as timers, Analog-to-Digital Converters (ADC), and communication modules---onto a single chip.
Compared to the microprocessors used in smartphones, tablets, and desktops, 
microcontrollers operate at lower frequencies and have smaller memory capacities.
Microcontrollers are found in a wide range of applications, including vehicles, medical devices, and industrial control systems.

However, designing and implementing secure microcontroller systems is challenging.
They often lack features that are standard in microprocessor architectures. 
For example, microcontrollers do not normally include a Memory Management Unit (MMU), which is used to implement privilege isolation, fine-grained memory access control, Address Space Layout Randomization (ASLR)~\cite{shi2022harm}, and many other security features.
Additionally, these systems are often programmed in low-level languages like C and assembly, which lack safety features and are prone to memory corruption bugs~\cite{szekeres2013sok, tan24sok, tan24canary}. 

Given their prevalence and use in critical applications, it is important to classify the challenges in designing and implementing secure microcontroller systems.
Unfortunately, the closed-source nature of most embedded and IoT systems presents substantial barriers to comprehensive analysis.
Even procuring real-world firmware samples is difficult, as highlighted in existing literature~\cite{wen2020firmxray, tan24sok, nino2024unveiling}.
Additionally, we argue that understanding and addressing these challenges requires more than just \emph{technical} analysis and solutions; it also involves grasping developer \emph{perceptions}.

For this study, we adopted a unique approach to study security challenges in microcontroller systems through the lens of the MITRE Embedded Capture the Flag (eCTF) competition~\cite{eCTFMitre}, an annual, months-long event.
Each year, participants are given a theme, an insecure reference system, and specific microcontrollers to develop upon.
Security is given top priority in the competition. 
Therefore, the security-related mistakes observed in eCTF are often fundamental, making them highly likely to appear in real-world development, where security is frequently not prioritized~\cite{venson2019impact, kreitz2019security}.
Indeed, some of our findings on vulnerabilities have been independently observed in research on real-world firmware analysis.

In contrast to existing studies that typically focus on a single aspect, such as firmware analysis~\cite{wen2020firmxray, tan24sok, nino2024unveiling}, and lack the capability to examine the broader perspective, our study takes a more comprehensive approach. 
Through the eCTF lens, we had the unique opportunity to explore the entire lifecycle of secure microcontroller system development --- from design documents and source code to binary analysis and developer perceptions. 
As a result, we not only confirm the presence of vulnerabilities but also gain first-hand insights into how these vulnerabilities emerge in the first place.
Furthermore, as the competitions attract early-stage firmware security researchers, our insights help enhance security education, prevent recurring issues, and guide future research.

Our study includes two sources of data: team submissions and post-competition interviews.
To make our submission analysis broad and thorough,
we attempted to identify and understand security-related mistakes and omissions made by teams.
We manually reviewed source code, documentation, and build tools, and we compiled every submission and examined relevant disassembly from the output.
We complemented the analysis with one-on-one interviews over Zoom.
Whereas the source code and documentation may tell us where mistakes exist, they cannot tell us why they exist.
Therefore, we used the interviews to gain a deeper understanding of participants' security acumen and to gauge whether or not they were aware of their mistakes and omissions.
This complementary approach proved to be a powerful tool for developing deep insights into which challenges are faced and why.

We break down our findings into two main categories.
In the first category, we detail the \emph{conceptual} challenges that participants faced.
These challenges are the result of a lack of knowledge or a misunderstanding.
In the second category, we detail the \emph{practical} challenges, which exist even when there is abundant knowledge available.
This dichotomy of results is useful because it allows the problem of securing microcontrollers to be approached from two sides.
Conceptually, we suggest that researchers, educators, and communities explore better ways to bridge the knowledge gaps faced by embedded system developers.
Practically, researchers and vendors should develop new methodologies and tools that not only identify and address security shortcomings but also lower the barrier for their deployment.

Our results highlight three main conceptual challenges and two main practical challenges in securing microcontroller systems.
Conceptually, there is a lack of knowledge about foundational security principles; it is difficult to adapt from a microprocessor system to a microcontroller one, and participants are not wholly aware of the unique attacks against microcontrollers or their defenses.
Practically, tools that naturally enhance security, like the memory-safe language Rust, lack sufficient support on microcontrollers, and we additionally find that the lack of high-quality entropy sources leads to less secure cryptography and secret generation.
The contributions of this paper are as follows:

\begin{itemize}
	\item We present an approach to studying the challenges in securing microcontroller systems through the lens of CTF competitions, which provides an opportunity to examine the entire lifecycle of the microcontroller system development from both technical and perceptual perspectives;
	\item By combining in-depth technical analysis with interviews, we uncovered both experiential and systemic security challenges, revealing key conceptual and practical difficulties in securing microcontroller system development;
	\item We offer actionable recommendations for researchers, developers, vendors, educators, and tool maintainers to address the identified challenges, bridge existing gaps, and strengthen the security of embedded systems.
\end{itemize}

\section{Background: MITRE eCTF Competition}

The eCTF is an annual, semester-long competition organized by MITRE where teams design, build, and attack ``secure'' embedded software for a given microcontroller platform.
Each competition has a topic, such as a firmware update system or a secure video game console~\cite{ectfposter}.
The competition consists of three phases: \textit{design/implementation}, \textit{handoff}, and \textit{attack}, and teams have just over four months to finish them.

In the design/implementation phase, teams are tasked with creating ``secure'' embedded software based on functional and security requirements.
Teams can use the provided reference design as a starting point or create their own design from scratch.
In the handoff phase, the event organizers verify that the functional requirements are met for the submitted source code.
Flags are then embedded into the firmware and must be protected by the defense mechanisms employed by the team.
In the attack phase, teams try to capture each other's flags by exploiting security weaknesses.

\textbf{Themes in 2023 and 2024 competitions}.
In the 2023 eCTF, teams were assigned the roles of car companies and were tasked with developing two sets of firmware for cars and key fobs with a remote keyless entry feature.
The firmware of the car and the fob runs on two development boards.
The 2024 eCTF focused on an insulin pump system consisting of one controller and two components: a blood sugar monitor and a pump actuator, all operating on three boards and communicating through an I\textsuperscript{2}C bus.

\textbf{Reference design}.
In both years, the organizers provided a reference design written in C as a starting point for competitors.
The reference design fulfills all the functional requirements but has no security features.
For example, all communication is in plaintext, with no defenses against hardware attacks, and several exploitable buffer overflow vulnerabilities exist in the provided functions.

\textbf{Microcontroller platform}.
The 2023 competition used a TI TM4C123GXL development board~\cite{tm4c123} equipped with two 80 MHz ARM Cortex-M4F microcontrollers.
The system has 256 KB of flash memory, 32 KB of SRAM, and a 2 KB EEPROM.
The 2024 competition used an Analog Devices Inc. MAX78000FTHR development board~\cite{max78000} equipped with a 100MHz ARM Cortex-M4 microcontroller and a 60MHz RISC-V coprocessor.
The board has 512KB of flash, 128KB of SRAM, and no EEPROM.

\textbf{Threat model}.
The threat model in the competition closely mirrors real-world scenarios involving embedded and IoT devices. 
The attacker is presumed to have physical access to the board and the communication channels between boards, enabling potential physical tampering besides software-based and network-based attacks. 
Additionally, the attacker has access to the firmware's source code. 
It is important to note, however, that the source code does not include any secrets or flags. 
These elements are generated and embedded into the protected firmware by the organizer, separate from the source code made available to the participants.

\section{Research Methodology}

We adopted a two-pronged approach: analysis of submissions and interviews with participants.
We included only teams that participated in the \emph{attack} phase, ensuring they had a functionally correct submission that represented their best effort to secure it.
Three PhD-level embedded-security researchers with prior eCTF experience conducted the submission and the interview data analysis collaboratively.

\subsection{Submission Analysis}

We analyzed 47 unique team submissions, with 20 from the 2023 competition and 27 from 2024, referring to individual teams as T1, T2, and so on.
This includes \textit{all} teams that passed functionality tests and entered the attack phase. The authors participated separately as teams \teama and \teamb, competing independently before collaborating to analyze findings and develop a taxonomy.
The authors' submissions were included in the statistical analysis for completeness but were excluded from examples and case studies to prevent bias.
The submissions consisted of source code, documentation, and build instructions, and we additionally had access to teams' posters~\cite{ectfposter} and presentations~\cite{ectf23presentation, ectf24ceremony}.
These artifacts were redistributed to all attack-phase participants, and every entrant was inherently given consent to their use for research purposes by signing the MITRE eCTF Participant Agreement~\cite{ectf24agreement}.
We additionally obtained written approval from MITRE for this study.

The criteria for submission analysis were informed by our experience with embedded security and participation in the eCTF over several years.
Through this experience, we have observed common security practices and where defenses and attacks typically occur in MCU-based systems, informing the selection of the following sections for review.

\textbf{Build tools}.
We examined the Makefiles, linker scripts, and documentation for each submission, and compiled them accordingly.
We noted the chosen programming language, compiler, and optimization level.
Additionally, we reviewed all security-related compiler flags and linker script attributes, as well as any warnings issued by the compiler during the compilation process.

\textbf{Source code and disassembly}. 
We applied a checklist-based manual review to the source code and disassembly of teams' submissions to gain insight into the defense mechanisms they employed and whether their behavior aligned with expectations.
The checklist was derived from prior experience and refined via post-competition team discussions and iterative analysis.
We used \texttt{git diff} to show changes from the reference files to ensure that we inspected all the code that was changed. 
In addition, we examined the disassembly of the compiled firmware to identify potential compiler optimizations that could negatively impact the system's security.

Specifically, we paid attention to particular keywords related to inline assembly, memory operations, random number generation, and timing in the source code and comments, such as \texttt{asm volatile}, \texttt{memset}, ``random'', and ``delay'', while examining the code.
We looked at random number generation because finding a reliable entropy source on a microcontroller can be challenging, and using an unreliable source can make the system less secure.
We investigated timing because adding random delays to a system in which an attacker has physical access can help reduce the efficacy of side-channel and fault injection attacks.
Additionally, we used debugging tools to gather runtime output for any uncertainty from the static analysis.

\subsection{Interviews with Participants}
\label{sec:interviews}

We conducted 22 semi-structured interviews with participants from 19 individual teams in the 2023 and 2024 eCTF competitions, with the exception of one who led another team between 2018 and 2021.
Among the interviewees, there were 12 undergraduate students, 6 Master's students, and 4 Ph.D. students.
Of these participants, 8 had never taken courses in computer security, while 14 had some experience in developing embedded systems.
No interviewees were from the authors' teams.

Table~\ref{t:demographic} provides demographic information of our participants.
For the participant ID convention, the first letter before the participant number indicates the year the participant competed (e.g., \textit{A} for 2023, \textit{B} for 2024, and \textit{C} for both 2023/2024 or multiple years), followed by the team number (e.g., \textit{A03-T11} for participant 03 in Team 11 who competed in 2023).

\begin{table}[!t]
	\centering
	\caption{Interview Participant Demographics.}
	\label{t:demographic}
	\setlength\tabcolsep{0.5ex}
	\small
	\renewcommand{\arraystretch}{1}
	\begin{tabular}{c|c|c|l|l|l|l|l} 
		\hline
		\multicolumn{1}{c|}{ID\textsuperscript{1}} & \multicolumn{2}{c|}{\begin{tabular}[c]{@{}c@{}}2023/\\2024\textsuperscript{2}\end{tabular}} & \multicolumn{1}{c|}{Major\textsuperscript{3}} & \begin{tabular}[c]{@{}l@{}}Edu\\Level\textsuperscript{4}\end{tabular} & \begin{tabular}[c]{@{}l@{}}CTF\\Pre-Exp\end{tabular} & \begin{tabular}[c]{@{}l@{}}Pre-Sec\\Courses\textsuperscript{5}\end{tabular} & \begin{tabular}[c]{@{}l@{}}Embed.\\Pre-Exp\end{tabular}  \\ 
		\hline\hline
		\parta      & \hspace{1.5px}Rust\hspace{1.5px} & Rust                                                                                                     & CS                                             & UG 2/3                                                                          & Very high                                            & Yes                                                                                      & None                                                     \\ 
		\hline
		\partb    & C    & C                                                                                                        & CS                                             & UG 2/3                                                                          & Very high                                            & Yes                                                                                      & None                                                     \\ 
		\hline
		\partc      & Rust & -                                                                                                        & CS,CY                                          & Master                                                                          & High                                                 & Yes                                                                                      & Limited                                                  \\ 
		\hline
		\partd    & C    & Rust                                                                                                     & CS                                             & Ph.D.                                                                           & High                                                 & Yes                                                                                      & None                                                     \\ 
		\hline
		\parte    & -    & C                                                                                                        & CS                                             & UG 3                                                                            & None                                                 & Yes                                                                                      & None                                                     \\ 
		\hline
		\partf    & -    & C                                                                                                        & CY                                             & UG 3                                                                            & Limited                                              & Yes                                                                                      & Limited                                                  \\ 
		\hline
		\partg    & C    & C                                                                                                        & CS,MA                                          & UG 1/2                                                                          & Medium                                               & No                                                                                       & None                                                     \\ 
		\hline
		\parth     & -    & C                                                                                                        & CS                                             & Master                                                                          & None                                                 & Yes                                                                                      & Limited                                                  \\ 
		\hline
		\parti    & C    & -                                                                                                        & CS                                             & UG 2                                                                            & Very high                                            & No                                                                                       & Limited                                                  \\ 
		\hline
		\partj & -    & C                                                                                                        & CE                                             & UG 1                                                                            & Limited                                              & No                                                                                       & None                                                     \\ 
		\hline
		\partk    & -    & C                                                                                                        & CLS                                            & UG 3/4                                                                          & None                                                 & No                                                                                       & Limited                                                  \\ 
		\hline
		\partl      & -    & C                                                                                                        & CY                                             & UG 4                                                                            & Limited                                              & Yes                                                                                      & Limited                                                  \\ 
		\hline
		\partm    & \multicolumn{2}{c|}{\textit{2018-2021}, C}                                                                      & CY                                             & MS/Ph.D.                                                                        & Very high                                            & Yes                                                                                      & High                                                     \\ 
		\hline
		\partn       & C    & -                                                                                                        & CY                                             & Ph.D.                                                                           & None                                                 & No                                                                                       & None                                                     \\ 
		\hline
		\parto    & -    & C                                                                                                        & CE,CY                                          & Master                                                                          & Limited                                              & Yes                                                                                      & Limited                                                  \\ 
		\hline
		\partp  & C    & -                                                                                                        & CS                                             & Ph.D.                                                                           & None                                                 & Yes                                                                                      & Limited                                                  \\ 
		\hline
		\partq     & -    & C                                                                                                        & CE                                             & Master                                                                          & None                                                 & No                                                                                       & High                                                     \\ 
		\hline
		\partr     & Rust & Rust                                                                                                     & CS                                             & UG 1/2                                                                          & Medium                                               & No                                                                                       & Limited                                                  \\ 
		\hline
		\parts      & -    & C                                                                                                        & CY                                             & Master                                                                          & Very high                                            & Yes                                                                                      & Limited                                                  \\ 
		\hline
		\partt   & -    & C                                                                                                        & CS                                             & UG 3                                                                            & None                                                 & No                                                                                       & None                                                     \\ 
		\hline
		\partu    & -    & C                                                                                                        & CE                                             & Master                                                                          & None                                                 & Yes                                                                                      & High                                                     \\ 
		\hline
		\partv   & -    & C                                                                                                        & CS                                             & UG 4                                                                            & Medium                                               & Yes                                                                                      & Limited                                                  \\
		\hline
	\end{tabular}
\begin{minipage}{8cm}
	\footnotesize
	\textbf{1}: The participant ID convention is described in Section~\ref{sec:interviews}.
	\textbf{2}: Indicates the programming language used by participant's team. ``-'' means the no attendance.
	\textbf{3}: CY: Cybersecurity; MA: Math; CE: Computer Engineering; CLS: Criminology, Law and Society.
	\textbf{4}: UG 2/3 means the participant was in undergraduate 2nd and 3rd year during 2023 and 2024 eCTF.
	\textbf{5}: Indicates whether participant had taken computer security related courses before the eCTF.
\end{minipage}
\end{table}

\textbf{Ethical considerations}.
We collaborated with the Institutional Review Boards (IRBs) at each of our institutions to ensure adherence to ethical guidelines, including informed consent, the right to withdrawal, and the anonymization of Personally Identifiable Information (PII) to protect participant privacy and confidentiality. 
After a thorough review, both of our IRBs determined that an IRB exemption was appropriate for this study.

\textbf{Participant recruitment}.
We conducted two rounds of participant recruitment to gather interviewees for our study.
Both rounds focused on attack phase participants who had competed at least once in the eCTF competitions from 2021 to 2024.
The first recruitment round took place at the 2024 eCTF award ceremony~\cite{ectf24ceremony} in April, while the second occurred in December via email outreach, with additional assistance from MITRE in promoting the study to past participants.
Note that while we could only recruit one pre-2023 participant, who led teams in 2018-2021 and advised in 2022-2023, their longitudinal perspective was valuable enough for inclusion in the study.
For all interested participants, we provided detailed study information and a consent information sheet outlining the study's purpose, the voluntary nature of participation, and the measures taken to ensure confidentiality and data security.
We recruited 10 participants in the first round and 12 in the second round, resulting in a total of 22 participants.

\textbf{Interview procedure}.
The interviews were conducted remotely via Zoom from May to June 2024 for the first round and from January to March 2025 for the second round.
They varied from 42 to 107 minutes, with an average of 69 minutes.
In addition to the consent process, participants were requested to fill out an online demographic survey form before the interview to streamline the interview process.
This form collected essential background information, such as their academic degree, area of study, and prior experience with security courses and competitions, particularly those relevant to CTF or embedded systems. 
Information regarding their role within their team and contributions during the competition was also gathered. 

The main body of the interview was driven by a pre-defined set of questions, focusing on the challenges participants faced, the strategies employed to secure their systems, and their understanding of security principles and mechanisms.
The interview questions were derived from the submission analysis to examine recurring issues.
Interviews were conducted in a one-to-one semi-structured way, with the option to opt out of any questions.
We compensate each interview participant with a \$50 gift card.

\textbf{Data collection and analysis}.
Data collection was conducted through recorded Zoom meetings.
The audio recordings were transcribed to text using Otter AI~\cite{otterai} services without PII.
We informed all participants about the privacy practices of Otter AI and obtained their consent to use the service.
To ensure the transcripts' fidelity, one of our study members reviewed and proofread each transcript against the audio recordings, correcting any discrepancies to preserve the semantic integrity of the participant responses.

Following transcription, the data analysis process began with the development of a preliminary coding scheme.
Guided by established qualitative and thematic analysis methods~\cite{saldana2021coding, clarke2014thematic}, three of our study members initially open-coded a single transcript independently to foster a diverse range of codes reflecting the intricacies of the interviews. 
These initial codes were then discussed collectively to formulate an agreed-upon codebook, following an inductive, reflexive workflow, which included categories that encapsulated the challenges faced by participants, the strategies they employed to address them, and any unique insights they shared.

This codebook guided the coding of subsequent transcripts, together with all three study members.
Weekly meetings were held to discuss the coding process, resolve any conflicts, and refine the codebook iteratively.
This collaborative approach negated the need for formal inter-rater reliability checks, as the codebook evolved through comprehensive consensus among the coders~\cite{mcdonald2019reliability}.
Finally, this resulted in a codebook consisting of 8 themes, 40 sub-themes, and 278 codes.
We make the permissible artifacts available\footnote{\url{https://github.com/CactiLab/eCTF-User-Study-Material}}.

\subsection{Stakeholders}

We focus on five stakeholder groups that are well-positioned to act on our findings and improve microcontroller software security:
(i) Researchers: academic/industrial scientists studying embedded security, 
(ii) Vendors: silicon, board, SDK, or RTOS providers, 
(iii) Educators: course or training designers, 
(iv) Developers: engineers who write or maintain MCU firmware, 
(v) Compiler maintainers: teams stewarding GCC/Clang and static-analysis pipelines.

\subsection{Threats to Validity}

In assessing the validity of our study, several limitations must be acknowledged. 

First, while the eCTF competition reflects many aspects of real-world embedded system security challenges, it remains a competition, and gamification elements such as point maximization may influence participant behavior in ways that deviate from real-world scenarios.
In addition, most eCTF teams are composed of students or early-career developers, which could constrain direct generalization to seasoned industrial settings; nonetheless, their challenges mirror the time-cost tradeoffs seen in industry, and prior studies of production firmware report similar missing defenses (§\ref{rq1-conceptual}), suggesting that our findings likely represent a conservative lower bound on real-world vulnerability prevalence.
As such, we do not claim that our conceptual and technical insights will fully generalize to other development environments.

Second, our focus on practical challenges in securing microcontroller systems meant that broader organizational aspects of the competition, such as team collaboration dynamics and the competition structure, were beyond the scope of our analysis.

Third, the categorization of challenges into conceptual and practical themes was intended to provide clarity and structure to the findings.
However, some sub-themes may overlap or extend beyond these categories, potentially introducing nuances not fully captured in the framework.
Additionally, while the study primarily centered on identified challenges, we included insights that were deemed beneficial to the community, which may have broadened the scope beyond the initial framework.

Fourth, the reliance on self-reported data collected through interviews introduces the possibility of social desirability bias.
Participants may have presented themselves or their teams in a more favorable light, which could affect the accuracy of the data.

Finally, the sample size (n=22), while typical for qualitative research, was relatively small and limited to participants from the eCTF competition.
Because we did not collect extensive data on participants' educational backgrounds, we cannot fully assess the adequacy of existing curricula, and our recommendations to educators should be viewed as preliminary suggestions rather than definitive conclusions.

\section{Conceptual Challenges}
\label{rq1-conceptual}

Conceptual challenges stem from gaps in knowledge or misunderstandings, and interviews are a highly effective tool for gaining insights into these issues.
Our analysis highlighted three key conceptual challenges in security:
security mechanisms, platform adaptation, and hardware attack and defense.

\subsection{Security Mechanisms}

Our study reveals three significant gaps in security mechanisms, namely privilege separation, memory wiping, and stack canary, which are basic but effective approaches for securing a system.
Figure~\ref{fig:analysis_features} shows the percentage of teams that implemented these security mechanisms in their designs.

\subsubsection{Privilege Separation}

Privilege separation is a security design principle that involves dividing a program or system into distinct components, each with different levels of privilege.
Cortex-M microcontrollers support privilege separation by offering hardware features such as the Memory Protection Unit (MPU) and distinct privileged and unprivileged execution modes.
The MPU can enforce read-only and non-executable memory and restrict access to configurable memory regions depending on execution mode.
These features must be enabled in code, either manually by the developer or through adequate support from the operating system.

\textbf{\textit{Submission Analysis:}}
We analyzed teams' submissions to understand whether and how they implemented privilege separation in their designs.
We found that even though privilege separation is a fundamental security concept, no teams in either year utilized it.
This is particularly notable given that this concept is frequently emphasized in security courses and widely recognized in the security community.
Moreover, recent studies have focused on making privilege separation easier to implement, more secure, and more efficient on microcontrollers.
For example, Kage~\cite{du2022holistic}, a compiler and FreeRTOS-based kernel, enhances control-flow protection by isolating the kernel.
Similarly, other research, such as Silhouette~\cite{zhou2020silhouette}, ACES~\cite{clements2018aces}, and EPOXY~\cite{clements2017protecting}, target bare-metal systems by lowering the privilege level of specific code segments.

\comparison{eCTF vs. real-world firmware}{
	Privilege separation is also rarely implemented in real-world microcontroller devices, appearing in only 1.78\% of firmware samples~\cite{tan24sok}. Understanding the reasons behind this is crucial.}

\textbf{\textit{Interviews:}}
Through our interviews, we sought to gauge participants' understanding of privilege separation, including whether they had any familiarity with it.
When participants were aware of privilege separation, we sought to understand why they didn't use it in their design.

\noindent\textbf{Many participants showed limited awareness and understanding of privilege separation on microcontrollers, often deterred by its perceived complexity.} 
While a few recognized the concept of least privilege, the intricacies of implementing it within the competition's timeframe or due to unfamiliarity with the microcontroller's low-level operations led to neglect in their design strategies.

19 out of 22 participants were unaware of the privilege separation feature in Cortex-M devices.
For instance, participant \parti understood the concept of least privilege but did not know that Cortex-M supports privilege separation.
They mentioned that even if they had become aware of it, they were still unsure if it was necessary for their design.
Similarly, participant \partm acknowledged that privilege separation could help in certain edge cases but argued that with this added complexity in design, \textit{``you will still potentially lose things.''}

For participants like \partb, even though they understood the concept of privilege separation, the low level at which microcontrollers are programmed caused them to overlook it: \textit{``[it] didn't even come into my mind.''} They continued:

\begin{quotationsm}
	\textit{``Because I don't know how the thing is working behind the scenes, I would just assume the level of privilege wouldn't work in my mind.''}
\end{quotationsm}

\noindent\textbf{Participants assumed that privilege separation is only useful in an OS-based environment and questioned its relevance in a bare-metal system.}
\parts and \partk were both unsure about the necessity of privilege separation in a bare-metal system.
\parts thought that privilege separation would be useful \textit{``if you have different processes and threads, different levels of execution while you're designing the system.''}
Since their system followed a flat, monolithic design, they did not see the need for privilege separation:

\begin{quotationsm}
	\textit{``It wasn't like we were trying to execute someone else's code in an unprivileged context, and protect it from the unprivileged one. So I don't think it was really ... useful.''}
\end{quotationsm}

\noindent\textbf{Participants believed that privilege separation would not further enhance their system's security.}
Participant \partl acknowledged that using both privilege levels is generally a good security practice, but did not identify any specific vulnerabilities in their design that would have been mitigated by it.
\parts's team prioritized avoiding implementation bugs over security mechanisms like privilege separation, assuming that a bug-free system would mitigate security risks:

\begin{quotationsm}
	\textit{``... the philosophy was, just don't have any bugs, and then you don't have to have any mitigations.''}
\end{quotationsm}

Similarly, \partd mentioned that while theoretically useful, privilege separation might not significantly enhance security in environments where \textit{``you have physical access already as the attacker.''}
They weighed the attributes of the system against the expected benefit of privilege separation, and concluded that privilege separation would not be worth the effort.

\noindent\textbf{Participants who had never heard of privilege separation tended to endorse it.}
While privilege separation can enhance security, its effectiveness is not guaranteed and heavily depends on careful implementation and system context. 
Introducing it may lead to compatibility challenges and potential new vulnerabilities if not properly managed.
Participant \partj said, \textit{``it probably would be a good way of making it [the system] more secure,''} and participant \parth was also inclined to favor it:
\begin{quotationsm}
	\textit{``I didn't know that existed. Yeah, that's really, really cool feature. We probably would have sent the team to go hunt for that if we had known existed.''}
\end{quotationsm}

\recommendation{
	\textbf{Researchers} should investigate barriers to privilege separation adoption, develop automated enforcement tools, and collaborate with educators and vendors to bridge theory and practice.
	\textbf{Vendors} should enhance their device support by providing demonstration projects that illustrate privilege separation or integrating suggestions into toolchains.
	We suggest that \textbf{educators} emphasize the importance of least privilege in system- and security-related courses, including specific strategies for implementing privilege separation in embedded systems.
}

\subsubsection{Memory Wiping} 
\label{sec:mem_wipe}

Cryptographic secrets or sensitive data, when stored in memory, pose a security risk if exposed. Their presence in memory increases vulnerability to unauthorized access through out-of-bounds reads~\cite{heartbleed} or cold boot attacks~\cite{halderman2009lest}.
Memory wiping is a technique used to minimize the duration that sensitive information remains in memory. 
However, memory wiping can fail if the vulnerability exists before the wipe, if it is not implemented properly, or if sensitive data is duplicated in memory. 
Nevertheless, memory wiping is an important part of good security hygiene.

\textbf{\textit{Submission Analysis:}}
We analyzed teams' submissions to understand whether, where, and how they used memory wiping.
Similar to privilege separation, memory wiping is foundational to security and provides significant value compared to the difficulty in implementing it.

\noindent\textbf{Teams attempted to use memory wiping in their designs, but their attempts were partially or fully nullified by the compiler.}
\teaml used \texttt{memset} three times in one function to zero out local buffers, as shown in Listing~\ref{code:teaml_memset}.
However, the analysis of their binary revealed that the second \texttt{memset}, which zeroes out the secret AES key from the stack, is optimized away by the compiler.
Indeed, \teaml used \texttt{memset} 10 times in their system, and 5 of them were optimized away.
Similar issues were found in \teamd and \teamj's submissions, in which the compiler optimized away 11 out of 30 and 2 out of 5 calls to \texttt{memset}, respectively.

\begin{listing}[!ht]
	\begin{minted}[xleftmargin=6pt, numbersep=5pt, tabsize=4, frame=lines, framesep=1mm, breaklines, linenos, fontsize=\footnotesize, escapeinside=||]{c}
void unlockCar(FLASH_DATA *fob_state_ram) {
	if (fob_state_ram->paired == FLASH_PAIRED) {
		...
		MESSAGE_PACKET message;
		char buffer[64];
		|\ul{memset}|(buffer, 0, 64);
		...
		message.buffer = buffer + 1;
		struct tc_aes_key_sched_struct s;
		...
		tc_aes_encrypt((message.buffer)-1, (message.buffer)-1, &s);
		...
		|\textbf{\textcolor{red}{\ul{memset}}}|(&s, 0, sizeof(struct tc_aes_key_sched_struct));
		|\ul{memset}|(message.buffer, 0, 64);
	}
}
	\end{minted}
	\vspace{-10pt}
	\caption{The C code of a function that uses \texttt{memset} to zeroize local buffers from \teaml. In the compiled firmware, the second \texttt{memset} was optimized away.}
	\label{code:teaml_memset}
\end{listing}

\noindent\textbf{Teams used library functions or custom inline functions to wipe memory, which effectively prevented the compiler from optimizing away the wipes.}
To effectively remove sensitive data from memory, \teamt and \teamo utilized the wiping function from the Monocypher library~\cite{monocypher} (Listing~\ref{code:crypto_wipe}), which uses the \texttt{volatile} keyword to prevent compiler optimizations.
For teams using Rust, both \teamn and \teamb utilized the Rust \texttt{zeroize} library to prevent optimization of buffer zeroization~\cite{rustzeroize}.

\begin{listing}[!ht]
	\begin{minted}[xleftmargin=6pt, numbersep=5pt, tabsize=4, frame=lines, framesep=1mm, breaklines, linenos, fontsize=\footnotesize, escapeinside=||]{c}
|\#|define ZERO(buf, size)  FOR(_i_, 0, size) (buf)[_i_] = 0

void crypto_wipe(void *secret, size_t size) {
	volatile u8 *v_secret = (u8*)secret;
	ZERO(v_secret, size);
}
	\end{minted}
	\vspace{-10pt}
	\caption{The \texttt{crypto\_wipe} function defined in the Monocypher library. The \texttt{volatile} keyword prevents potential compiler optimizations.}
	\label{code:crypto_wipe}
\end{listing}

\teamr implemented an inline function to erase sensitive data on the stack, as shown in Listing~\ref{code:teamr_wipe}.
After examining the resulting firmware, we observed that the compiler either unrolled the calls to this inline function for small \texttt{data\_len} values or replaced them with \texttt{memset} calls for larger \texttt{data\_len} values, while none of the calls were optimized away.

\begin{listing}[!h]
	\begin{minted}[xleftmargin=6pt, numbersep=5pt, tabsize=4, frame=lines, framesep=1mm, breaklines, linenos, fontsize=\footnotesize, escapeinside=||]{c}
inline __attribute__((__always_inline__)) void erase_stack_data(uint8_t *start_add, uint32_t data_len) {
	for(uint32_t i=0; i<data_len; i++) start_add[i]=0;
}
	\end{minted}
	\vspace{-10pt}
	\caption{The inline function to erase used data on stack implemented by \teamr. }
	\label{code:teamr_wipe}
\end{listing}

\observation{
	Correctly implemented compiler optimizations cannot preserve the security-related program states that exceed the scope of semantic functionalities of language specifications~\cite{xu2023silent}.
	This means the security-related operations need to be explicitly controlled and verified by the developer.
}

\textbf{\textit{Interviews:}}
While analyzing the submissions and firmware was informative, it did not tell us whether participants were aware that the compiler could alter their wiping.
Therefore, our interviews sought to gauge participants' understanding of memory wiping and their awareness of potential compiler alterations to the code.

\noindent\textbf{Participants thought it was less effective to implement memory wiping because of the embedded systems' threat model.}
Participant \parta explains that in their threat model, the potential of an attacker gaining arbitrary memory read capabilities would allow access to firmware directly since they are all mapped in the same address space, which will negate the benefits of wiping memory at the application level.
\partl also mentioned that they designed their system under the assumption that memory could be dumped.
In addition, participant \partc said, \textit{``And if they found a way to do that [read memory content], they could probably do a lot worse than just reading intermediate memory like that.''}

\noindent\textbf{Participants who used the standard library functions to erase the memory were not aware that the compiler could optimize them away.}
Participant \partg's and \partk's teams utilized \texttt{memset} at the end of functions to erase leftover content.
However, they were not aware that the compiler was optimizing away their \texttt{memset} calls.
As participant \partg realized:

\begin{quotationsm}
	\textit{``That's interesting. I didn't think that it would optimize that out. Was that because of the optimization flags?''}
\end{quotationsm}

\recommendation{
	\textbf{Developers} should use vetted zero-memory primitives (e.g., \texttt{memzero\_explicit}) or wiping functions from the trusted cryptographic libraries to guarantee non-elision.
	\textbf{Researchers} should design and evaluate annotation schemes that compilers can honor to preserve security-critical wipes, accompanied by automated tooling for verification.
	\textbf{Compiler} should notify users when code that has potential security implications is optimized away and provide suggestions for alternatives.
}

\subsubsection{Stack Canary} 
\label{sec:stack_canary}

Stack canary is a defensive mechanism designed to detect and mitigate buffer overflow attacks by inserting a known, random value---referred to as a canary---into the stack frame just before the return address.
At the conclusion of a function, the integrity of this canary is verified; any modification suggests that a buffer overflow has occurred, prompting the system to take protective measures such as halting execution or invoking an exception handler.
On Cortex-M microcontroller systems, the stack canary feature can be enabled by configuring compiler-level protections.
Modern toolchains, such as GCC, support this mechanism through options like \texttt{-fstack-protector} or \texttt{-fstack-protector-strong}, which automatically instrument code with canary checks.

\textbf{\textit{Submission Analysis:}}
We analyzed teams' submissions to understand whether and how they implemented stack canaries.

\noindent\textbf{Among the 47 teams, only 2 (4.26\%) enabled stack canary protection by activating the appropriate compiler flags.}
Notably, none of these teams provided additional initialization to randomize the canary value or to customize the error handler.
In the absence of user-supplied initialization, the toolchain library defaults to defining the canary as a fixed constant.
Consequently, if a buffer overflow corrupts the canary, the default weak-defined handler---typically designed to halt execution or trigger a system reset---will be invoked unless it is explicitly overridden by the user~\cite{newlibssp, tan24canary}.
Despite this default configuration, incorporating the stack canary remains a significant step towards securing against buffer overflow attacks.

\comparison{eCTF vs. real-world firmware}{
	Stack canaries are rarely implemented in real-world microcontroller devices, with a presence rate below 0.2\% in large-scale firmware samples~\cite{tan24sok, nino2024unveiling}. Even when implemented, Xi et al.~\cite{tan24canary} found them less effective due to the lack of canary randomization and prolonged reuse.
}

\textbf{\textit{Interviews:}}
In our interviews, we aimed to understand why participants did not implement stack canaries, as well as their perceptions of the potential benefits or limitations of using stack canaries in microcontroller systems.

\noindent\textbf{Participants who understood but did not implement stack canaries thought they needed to be enabled by manually inserting the canary instructions.}
They were not aware that the stack canary feature could be enabled by configuring compiler flags.
Participant \parts incorrectly believed that stack canaries are not typically available on embedded systems, and expressed interest in implementing them manually:

\begin{quotationsm}
	\textit{``... on embedded systems that this [stack canary] is not a feature that is usually present ... it would have been fun to do a little implementation of a stack canary, make like an LLVM pass that will automatically inject ... some instructions that will do that for us.''}
\end{quotationsm}

Similarly, \partu manually implemented a basic stack canary mechanism by writing a hardcoded pattern (`A's) above the stack to detect overflows.
They were unaware that the compiler could automatically insert stack canaries with the appropriate flags.

\noindent\textbf{Participants were not aware that the stack canary feature could be less effective on microcontroller systems.}
Tan et al.~\cite{tan24canary} found that the stack canary feature is less effective on microcontroller systems due to the lack of system support for randomizing the canary value and the prolonged reuse of a single canary value.
Among all participants, only \partm were aware that stack canaries on embedded systems can sometimes be static, making them vulnerable to being leaked through memory dumps or crashes.
They noted that an attacker with sufficient knowledge could determine the canary value and bypass its protection:

\begin{quotationsm}
	\textit{``But realistically, if you have someone who understands how things work, I mean, you figure out what the stack canary is, and you just use that [to bypass it].''}
\end{quotationsm}

\recommendation{
	\textbf{Researchers} should design more effective stack canary mechanisms for microcontroller systems.
	\textbf{Vendors} should provide options to enable stack canaries when building with their toolchains.
}

\begin{figure}[t]
	\centering
	\includegraphics[width=\linewidth]{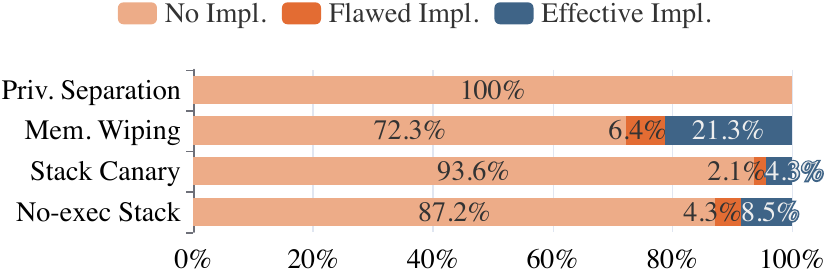}
	\caption{The security mechanisms implemented by the teams in all 47 designs over both years.}
	\Description{The security mechanisms implemented by the teams in all 47 designs over both years.}
	\label{fig:analysis_features}
\end{figure}

\subsection{Platform Adaptation}

System developers must be aware of the platform they're working on, especially if they're moving from one platform to another. For example, a developer writing secure software for a microcontroller must be aware that their hardware likely does not include an MMU. 

By analyzing teams' submissions and interviewing participants, we sought to understand the conceptual challenges that developers face when moving from microprocessor systems to microcontroller systems. Our analysis revealed two key areas that are related to memory access control, in which developers had challenges to adapt: non-executable stack and relocation read-only.

\subsubsection{Non-executable Stack}
\label{sec:noexecstack}
Making the stack non-executable can effectively thwart stack-based code injection attacks.
In modern microprocessor systems (e.g., Cortex-A or x86/64), ensuring that the stack of applications is non-executable involves two major steps.
First, during the compilation and linking stage, developers specify that they want the stack to be non-executable by using specific linking options. 
The latest versions of GCC and LLVM default to using a non-executable stack by setting attributes in the program header of the ELF binary~\cite{execstack}.
Second, when the program is loaded into memory, the loader reads the ELF's program header and asks the operating system to enforce non-executability by configuring the MMU page settings.

However, the MCU system does not load the program during runtime; instead, the raw binary, without the program header, eventually gets copied out of the ELF binary and flashed to the MCU's storage media.
Thus, the memory attributes in the ELF's program header are already lost during this process.
As a result, teams that aimed to enforce non-executability had to devise their own techniques, utilizing the MPU to protect access to stack memory.

\textbf{\textit{Submission Analysis:}}
We sought to understand whether and how they attempted to make the stack non-executable without an operating system.
We believe that this knowledge is crucial because preventing the execution of data on the stack is another hugely effective foundational security concept.

\noindent\textbf{Only four submissions successfully implemented stack non-executability.}
As shown in Figure~\ref{fig:analysis_features}, four teams out of 47 across both years manually configured the MPU during the firmware initialization process to make the stack non-executable.
This entailed enabling the MPU, configuring the memory region for the stack with the eXecute Never (XN) attribute set, and subsequently enabling the region.

\noindent\textbf{A few teams tried to enable the non-executable stack but failed to do so.}
Two other teams attempted to make the stack non-executable, but they only completed the first step of marking the ELF header and missed the second step of honoring the request during the firmware initialization process.
For instance, \teamr modified the attribute of the whole SRAM to non-executable in their linker script (Listing~\ref{code:soft_linker}).
\teamm also explicitly specified the linker option \texttt{-z noexecstack} during the compilation of the ELF binary.

\begin{listing}[!ht]
	\begin{minted}[xleftmargin=6pt, numbersep=5pt, tabsize=4, frame=lines, framesep=1mm, breaklines, linenos, fontsize=\footnotesize, escapeinside=||]{text}
MEMORY
{
	FLASH (rx) : ORIGIN = 0x00008000, LENGTH = 0x00038000
	SRAM  (|\textbf{\textcolor{purple}{rw}}|) : ORIGIN = 0x20000000, LENGTH = 0x00008000
}
	\end{minted}
	\vspace{-10pt}
	\caption{The \teamr's linker script snippet showing that the SRAM region is set to read and write only.}
	\label{code:soft_linker}
\end{listing}

\comparison{eCTF vs. real-world firmware}{
	The real-world usage of MPU for memory protection is also minimal, with presence rates less than 2\% in large-scale firmware samples~\cite{tan24sok, nino2024unveiling}.
}

\textbf{\textit{Interviews:}}
Through our interviews, we sought to understand why participants didn't attempt to implement non-executability, and if they did, whether they knew that their implementation did not work and why it did not work.

\noindent\textbf{More than half of the participants (12/22) were unaware of the advantages of a non-executable stack and how to implement it.}
Participant \parta was not aware of the non-executable stack feature during the competition.
As their team used Rust, they were also unsure \textit{``if using the Rust compiler specifically for an embedded target will also set those memory protection flags correctly.''}
\partb mentioned that they were \textit{``only looking at library specific flags''} when implementing the crypto, and similarly \partl admitted that \textit{``the compiler was a potential that we left on the table.''}

\partg noticed some teams configured the memory pages in C code and then \textit{``setting the bits''} on them after the competition, but they did not find out the specific reasons for doing so.

\noindent\textbf{Most participants (21/22) thought adding the \texttt{noexecstack} compiler flag or modifying the attributes in the linker script would effectively make the stack non-executable on a microcontroller system.}
Participant \parte believed that the non-executable stack flag is a standard method to increase security against buffer overflow exploits in microcontroller environments:

\begin{quotationsm}
	\textit{``Yeah, so I think like, non executable stack is something very basic ... very much like standard and a lot more protective in terms of making it difficult to [exploit].''}
\end{quotationsm}

Participant \partd thought it would be a good but incomplete defense when compiling the embedded binary with the non-executable stack:

\begin{quotationsm}
	\textit{``So it's like they're definitely, obviously it's like not a complete solution. But ... I would imagine that would still be useful as well in an embedded scenario.''}
\end{quotationsm}

Participant \parti endorsed the idea of enabling the non-executable stack flag and thought it may work similarly to a traditional microprocessor system:

\begin{quotationsm}
	\textit{``It enables the memory protections, I think, for that region of virtual memory? As to how I work on the microcontroller, probably similar to ... a normal PC.''}
\end{quotationsm}

Only participant \partc acknowledged that merely setting the relevant bits in the ELF files will not make the RAM region non-executable on a microcontroller system.

\recommendation{
	\textbf{Researchers} should explore compiler or linker extensions that automatically enforce memory attributes in microcontroller systems.
	\textbf{Vendors} might consider preserving the necessary attributes to the raw binary that is flashed to the device in their build toolchains, which enables firmware to read them during initialization to configure corresponding permissions.
}

\subsubsection{Relocation Read-only (RELRO)}
In dynamically linked ELF binaries, the Global Offset Table (GOT) stores function pointers that are resolved during dynamic linking. 
Overwriting these function pointers has been an effective attack vector for control-flow hijacking.
RELRO~\cite{relro} is a binary hardening technique to mark GOT and related sections as read-only in the ELF files to mitigate control-flow hijacking.
Similarly to enforcing the non-executable stack, it's imperative for both the loader and the operating system to recognize and adhere to this marking, configuring the memory settings accordingly.
However, microcontroller firmware is usually statically linked, which makes the presence of GOT and related sections uncommon, rendering RELRO ineffective for such systems.

\textbf{\textit{Submission Analysis:}}
\teamm enabled the \texttt{-z relro} linker flag in their Makefile. However, since their firmware is statically linked, this flag has no effect.
Unfortunately, we did not have a chance to interview them about their choice.
Note that the compiler also does not give warnings when potential invalid options for a specific architecture have been enabled.

\recommendation{
	\textbf{Compiler} should warn users when they enable options that may not work on the target architecture. They should also refer users to their architecture's alternatives that keep the same security properties when applicable.
}

\subsection{Hardware Attack and Defense}
\label{sec:attdef}

Compared to microprocessor systems, MCU-based embedded devices are more susceptible to hardware attacks such as tampering with the physical communication channel, side-channel analysis, and fault injection due to their physical accessibility and simpler circuit design.
We refer to them as \textit{embedded-prone attacks}.
As a result, developers must practice good security hygiene while additionally accounting for these unique threats.

\begin{figure}[t]
	\centering
	\includegraphics[width=\linewidth]{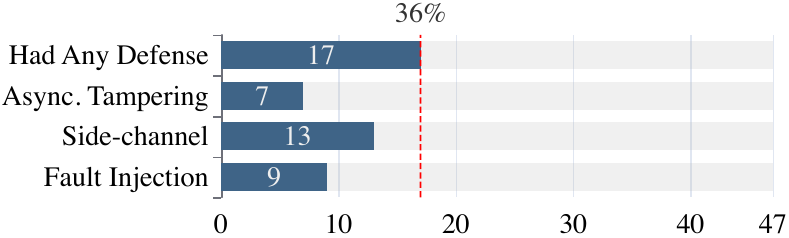}
	\caption{The defense applied by teams against embedded-prone attacks in both years.}
	\Description{The defense applied by teams against embedded-prone attacks in both years.}
	\label{fig:analysis_hardware}
\end{figure}

\textbf{\textit{Submission Analysis:}}
We analyzed teams' submissions to understand whether and how defenses were implemented against embedded-prone attacks.
We are careful not to include defenses against attacks that are equally practicable on the microprocessor system, as they were not the focus of this study.
The analysis results are shown in Figure~\ref{fig:analysis_hardware}.

\noindent\textbf{Across both years, only 17/47 teams (36.17\%) implemented any defense against embedded-prone attacks.}
Rather, the majority of competitors focused on securing their communication and defending against common software vulnerabilities.
For example, most competitors were aware of buffer overflow and brute-force vulnerabilities, and chose to avoid using unsafe functions like \texttt{gets} as a result.

\noindent\textbf{Of the teams that implemented any defenses, asynchronous physical communication channel tampering was the least frequently defended against, at 7/17 (41.18\%).}
Asynchronous means the attacker can intercept the communication for offline tampering for as long as they want.
For example, \teamo designed its communication protocol to require a timely response from the other side.
They implemented a timeout mechanism such that messages must have been received within a certain time frame to be accepted as valid.
If attackers want to manipulate the message in the physical channel, they need to modify the intercepted messages in a tiny time window, which might not be feasible.

\noindent\textbf{Of the teams that implemented any defenses, side-channel analysis was the most frequently defended against, at 13/17 (76.47\%).}
These defenses effectively prevent timing side-channel analysis on password comparison by completing the process in constant time regardless of the input.
For example, \teamg used the \texttt{timing\_safe\_strcmp} function from the bCrypt~\cite{bcrypt} library, while teams that did not implement a defense used the \texttt{strcmp()} or \texttt{memcmp()} library functions, which do not operate in constant-time.
Other teams, such as \teamu, implemented their own constant-time comparison function with logical operators (Listing~\ref{code:teamu_compare}).
Given a fixed \texttt{length} argument, the execution time of this function is constant and is independent of the similarity of the two string inputs.

\begin{listing}[t]
	\begin{minted}[xleftmargin=6pt, numbersep=5pt, tabsize=4, frame=lines, framesep=1mm, breaklines, linenos, fontsize=\footnotesize]{c}
int ConstCompare(const uint8_t* a, const uint8_t* b, int length) {
	int i;
	int compareSum = 0;
	for (i = 0; i < length; i++) {
		compareSum |= a[i] ^ b[i];
	}
	return compareSum;
}
	\end{minted}
	\vspace{-10pt}
	\caption{The constant-time password comparison function from \teamu.}
	\label{code:teamu_compare}
\end{listing}

\noindent\textbf{Of the teams that implemented any defenses, fault injection attacks were defended against by 9/17 teams (52.94\%).}
Mitigation inserts random delays to make it harder for an attacker to inject a fault at the correct time.
\teamo implemented the random delay macro to delay 1 to 255 CPU cycles and check for possible glitches at the end (Listing~\ref{code:teamo_delay}).
Using the macro can also avoid potential unwanted compiler optimizations~\cite{macroinc}.

\begin{listing}[!h]
	\begin{minted}[xleftmargin=6pt, numbersep=5pt, tabsize=4, frame=lines, framesep=1mm, breaklines, linenos, fontsize=\footnotesize, escapeinside=||]{c}
|\#|define RAND_STALL() \
	rand_ret = -1; \
	rand_ret = fillEntropyBuf(rand_rbt, 2); \
	if (rand_ret == -1) halt_and_catch_fire(); \
	rand_i = 0; \
	rand_y = 0; \
	for (rand_i = 0; rand_i < rand_rbt[0]; rand_i++) \
		rand_y += 1; \
	rand_ret = ((rand_i == rand_y) && (rand_rbt[0] == rand_y))
	\end{minted}
	\vspace{-10pt}
	\caption{The random delay macro from \teamo.}
	\label{code:teamo_delay}
\end{listing}

\textbf{\textit{Interviews:}}
We sought to gauge participants' knowledge of the unique attack surface in the microcontroller environment and understand why they either did not implement defenses against embedded-prone attacks or did not implement all possible defenses.

\noindent\textbf{Participants acknowledged the larger attack surface of embedded systems due to physical accessibility.}
As mentioned by participant \partd, \textit{``we had to assume that attackers had physical access with the board,''} which influenced their defensive strategies.
They need to additionally consider physical attacks, such as side-channel, and assume the attacker can do \textit{``a lot of things that cryptography alone isn't really equipped to handle.''}

\partk believed that the embedded systems are simpler but easier to attack because \textit{``it does not have an OS, and everything is pretty much streamlined, so you do have a lot of good regularities in that systems.''}
In addition, they are more susceptible to physical attacks as \textit{``you are exposing to the hardware immediately.''}

\noindent\textbf{Participants gave less priority to defending against hardware attacks than software vulnerabilities.}
Participant \parth initially aimed to secure the design against side-channel analysis.
However, due to their knowledge gap in the side-channel analysis, and \textit{``we realized that the number of people who actually know how to run these attacks is actually quite small,''} they decided to focus on preventing buffer overflows instead.

Participant \parti admitted that the fault injection attacks were \textit{``more difficult compared to some of the low hanging fruit,''} and they lacked awareness of straightforward protective strategies.
As a result, they instead focused on defending attacks that they thought \textit{``would be easier to execute.''}
And participant \partv also commented on the hardware attacks as they \textit{``can be much difficult to protect from.''}

\partm admitted that it is \textit{``technically easier''} to conduct hardware attacks on embedded devices, and attributed the lower priority of hardware defenses to the knowledge gap among developers:

\begin{quotationsm}
	\textit{``It's harder to do hardware stuff because you're missing the knowledge. It's easier to do software because more people ... understand how software attacks work.''}
\end{quotationsm}

Participant \parts felt that software security is generally easier to implement and provides higher defensive value compared to hardware-based protections:

\begin{quotationsm}
	\textit{``The potential upsides of defending from a software perspective are both higher and cheaper for ... developer or manufacturer than on the hardware side.''}
\end{quotationsm}

\observation{
	The knowledge gap in embedded-prone attacks was the dominant factor that prevented participants from implementing effective defenses.
}

\noindent\textbf{Even if participants were aware of embedded-prone hardware attacks, they had challenges implementing defenses.}
Participant \partb discussed the challenges of designing persistent memory on microcontrollers due to the paged flash access:

\begin{quotationsm}
	\textit{``We had to find pages in memory, that we're not going to overwrite anything else that was useful, which usually in general-purpose, that is done for you ... It's already there in the logistics of the system.''}
\end{quotationsm}

Participant \parto admitted that they \textit{``don't know if there is a defend mechanism against glitching''} and thus skipped the defense.
Similarly, \parth acknowledged the theoretical possibility of fault injection attacks but lacked the resources and tools to effectively explore and mitigate such threats:

\begin{quotationsm}
	\textit{``For the CS guys, we had no clue this [fault injection] was a thing. And even for the embedded systems [CE] guys, they were like, we know this theoretically could be done, but we don't quite know how to do it. And we don't know how to prevent against it.''}
\end{quotationsm}

Additionally, besides being harder to execute, \partr believed that the hardware attacks are \textit{``more difficult to mitigate''} than software attacks.
They also noted that compilers provide little to no protection against hardware attacks and may even \textit{``work against you''} by introducing unintended security weaknesses.

\recommendation{
	\textbf{Researchers} should explore compiler or build-system approaches for seamless integration of hardware protections.
	\textbf{Vendors} can improve hardware or library support for logging anomalies and detecting tampering events at runtime.
	We suggest that \textbf{educators} incorporate hardware attack scenarios into security curricula.
	Hands-on labs and simplified tools can help students both apply hardware defenses and learn to prioritize protections under realistic constraints.
}

\section{Practical Challenges} 
\label{rq2-practical}

Practical challenges are those faced even when there is an abundance of information.
Based on our analysis, two practical challenges emerged as particularly significant: embedded Rust adaptation and sources of entropy.

\subsection{Memory Safe Language} \label{sec:mem_safe_prog}
Microcontroller systems require permissive access to memory to properly and efficiently interface with hardware and peripherals.
Accordingly, they are usually programmed in C, but the difficulty of safely and correctly accessing memory often leads to vulnerabilities.

Rust, a memory-safe language, aims to eliminate or reduce memory corruption vulnerabilities with compile-time safety checks.
Because Rust is checked for issues at compile time, it maintains a speed similar to that of C.
Unfortunately, the unique requirements of microcontroller systems present practical challenges to adopting Rust, which is shown in our interviews and submission analysis.

\textbf{\textit{Submission Analysis:}}
In 2023, 5 out of 20 teams that advanced to the attack phase used Rust as their primary programming language, whereas in 2024, only 2 out of 27 teams adopted Rust in their design. 
One contributing factor to this decline was that in 2024, the vendor-provided C Software Development Kit (SDK) was considerably larger in size compared to 2023.
Because most teams adopting Rust still needed to compile the vendor's C SDK with their Rust code, this led to excessive binary sizes that often exceeded flash memory limits.
However, as surfaced in our interviews, other adoption barriers also influenced teams' decisions to use Rust.

\textbf{\textit{Interviews:}}
We prepared interview questions to understand participants' views on using a memory-safe language within the microcontroller environment and to learn about their attitudes and practices regarding code safety.

\subsubsection{Perspectives and Hesitations About Employing Rust}
While most participants did not choose Rust accounting for the overall team's familiarity, several participants described practical factors that limited or dissuaded them from incorporating Rust.

\observation{
	Besides familiarity, participants decided against Rust for reasons spanning the lack of direct vendor support, insufficient library support for microcontroller systems, and the steep learning curve.
	Participants who did not use Rust also expressed misconceptions about its memory safety features.
}

\noindent\textbf{Participants cited the lack of direct vendor support and insufficient library support for microcontroller systems as major obstacles to switching to Rust.}
For both years, the SDK for interfacing with the hardware and peripherals was only provided in C. To use Rust, teams had to write their own implementations of SDK functions or use Rust-to-C bindings to interact with hardware.

\parte, for instance, realized that the \textit{``[vendor] library support is not present in Rust,''} which made hardware interactions, such as I2C or flash memory, exceedingly difficult without rewriting entire driver layers from scratch.

Participant \partm's team also focused on language compatibility with the target architecture, explaining that having a functional compiler and toolchain was a deciding factor for language choice:

\begin{quotationsm}
	\textit{``Okay, is there a compiler for that particular architecture which allows us to compile binary? And the other thing was, like, you don't need only a compiler, but you need also, like the whole toolchain.''}
\end{quotationsm}

They believed Rust was not widely adopted in commercial embedded development due to the extra effort required in the toolchain:

\begin{quotationsm}
	\textit{``I have something like 800 [market] embedded IoT devices ... No one in their right mind would use Rust.''}
\end{quotationsm}

\noindent\textbf{The steep learning curve was also a major hurdle for those who hesitated to adopt Rust.}
\partd recognized Rust's learning curve as a significant time investment, especially nuances like handling unsafe code, which were unfamiliar to those experienced in C/C++:

\begin{quotationsm}
	\textit{``... especially working with unsafe code and stuff, there's a lot of ... weird idiosyncrasies with Rust that I think what it slowed us down in the development.''}
\end{quotationsm}

\partr found Rust more complex than C due to features like the ownership model, which \textit{``people have to understand''} to take its full advantage.
Participant \parti acknowledged the advantages of Rust in terms of security and robustness but expressed concerns that it \textit{``would be more difficult to develop rapidly.''}

\noindent\textbf{Participants who had not actively used Rust sometimes saw it as \emph{completely eliminating} memory vulnerabilities.}
Seven participants who did not use Rust expressed misconceptions about its memory safety features on embedded systems.
Participant \partj said that in their understanding, Rust takes care of memory safety; it is \textit{``already memory safe''} compared to other languages.
\partl also thought that Rust by nature \textit{``doesn't let you write the function if it's not memory safe, or it kind of optimizes it out if it's not memory safe.''}
While it is generally true that Rust enforces strong safety guarantees, these participants did not realize the challenges posed by \texttt{unsafe} code blocks or low-level memory interactions, which are discussed in the next section.

\recommendation{
	To bridge the knowledge gap, the researchers, vendors, and educators within the \textbf{Rust community} might consider emphasizing that while Rust ensures memory safety, challenges remain in microcontroller systems, especially when interfacing C libraries with \texttt{unsafe} blocks and involving low-level memory interactions.
}

\subsubsection{Adoption Experiences and Challenges Among Rust Users}
In contrast, some participants did integrate Rust into their designs and encountered a different set of hurdles. 

\observation{
	Participants who used Rust mainly faced challenges in compiling Rust without the standard library, efficiently implementing the hardware abstraction layer, and managing \texttt{unsafe} operations properly.
}

\noindent\textbf{Participants encountered challenges when compiling Rust without the standard library on the embedded device.}
The Rust standard library provides abstractions, types, operations, and other quality-of-life utilities for developing in Rust~\cite{ruststd}. 
However, it is often necessary to use Rust without the standard environment in situations where an OS is not available.

\partc told us that transitioning to using Rust in a \texttt{no\_std}~\cite{rustnostd} environment posed a unique set of challenges distinct from general-purpose programming on the microprocessor systems.
They had to ensure that any dependencies they wanted to use were compatible with a \texttt{no\_std} environment, which narrowed the range of usable libraries.
Besides compatibility, participant \partd also noted that any inclusion of the Rust standard library risked bloating the binary.

\noindent\textbf{Participants highlighted the challenges in implementing a secure Hardware Abstraction Layer (HAL).}
A common practice in embedded development is to perform hardware interactions through an intermediate HAL, which hides the low-level implementation details from the rest of the system. 
The HAL used for the reference system was provided as an SDK that offered C interfaces to various components of the microcontroller used in the competition.
However, a similar HAL was not available for Rust.

Participant \parta told us that in 2023, they created many Rust bindings to the original C SDK, which was compiled with their Rust code that only implements the high-level protocol:

\begin{quotationsm}
	\textit{``... if you look at our 2023 codebase, it's mostly ... written in C because we just collect the entire C library for the Tiva [C SDK]. That's instruments driver, and then we compile that in with our like, small Rust code. ''}
\end{quotationsm}

In the 2024 event, \parta's team avoided compiling the vendor's large C SDK alongside their Rust code (which inflated firmware size), and instead wrote a minimal HAL from scratch:

\begin{quotationsm}
	\textit{``... it lets us have a more holistic view of the device ... It lets us get more Rust experience and make sure that we're also avoiding any bugs that could potentially occur in a C library.''}
\end{quotationsm}

In order to develop the HAL that directly interacts with peripherals, \parta's team utilized the svd2rust~\cite{rustsvd2rust} tool to automatically generate Rust structures from CMSIS-SVD~\cite{armcmsis} files, which describe the memory-mapped registers of peripherals and are available in the vendor-provided SDKs. 
This automation reduced the need for manual volatile reads/writes by providing pre-generated bindings for hardware interactions.
After this, with the Rust peripheral access crate~\cite{rustpac}, they were able to abstract the direct register manipulations into more manageable function calls in HAL. 

\observation{
	Automated conversion of hardware descriptor files into Rust structures can accelerate embedded HAL development, but its fidelity and security require further research.
}

\parta also mentioned that they encountered inconsistencies between the vendor-provided CMSIS-SVD files and the device user manual.
They believe \textit{``it's probably because they [the vendor] took some previous SVD file copied over and change.''}

\noindent\textbf{Teams also struggled with effectively compartmentalizing \texttt{unsafe} operations.}
It is possible to call C functions from Rust code, but only within an unsafe block. 
Unsafe blocks in Rust allow for certain operations that the compiler cannot guarantee to be safe, such as dereferencing raw pointers or calling external C code~\cite{rustunsafe}.
This bypasses the rigorous safety checks normally enforced by Rust, potentially leading to security vulnerabilities such as buffer overflows or access violations if not carefully managed.

Some teams, such as \teamk led by \partc, used \texttt{unsafe} blocks extensively for calling C SDK functions, primarily due to their incremental development model, which allowed for parts of the system to be gradually ported to Rust while continuously testing and validating functionality.
They explained:
\begin{quotationsm}
	\textit{``... the teams that chose to rewrite everything in scratch from Rust would have had an intermediate state, your part of the code was ported over to Rust, part of it was in C, but they were separate. And there was no way to know if it worked until the Rust rewrite was done.''}
\end{quotationsm}

In contrast, \partr's team aimed to confine \texttt{unsafe} code to low-level crates that interact with hardware, while they tried not to use unsafe code \textit{``especially like anywhere in our logic.''}

\recommendation{
	To ease the secure HAL development, \textbf{researchers} should further explore ways to automate the conversion of hardware descriptor files into Rust structures and devise methods to identify the insecure use of \texttt{unsafe} blocks.
	\textbf{Vendors} should consider enhancing their support for Rust in embedded systems by providing full Rust SDKs or Rust-to-C bindings for their existing C libraries to ease the transition for developers.
}

\subsection{Entropy Sources}

Like microprocessor systems, microcontrollers use Pseudo-Random Number Generators (PRNGs) to generate cryptographic secrets.
PRNGs take a seed as an input to deterministically generate an output.
Since the output is deterministic, the input must be comprised of harvested randomness.
However, there are fewer high-quality sources of entropy on a microcontroller system than a microprocessor system, complicating the generation of pseudo-random numbers and impacting the robustness of cryptographic operations~\cite{heninger2012mining}.

\begin{figure}[t]
	\centering
	\includegraphics[width=0.98\linewidth]{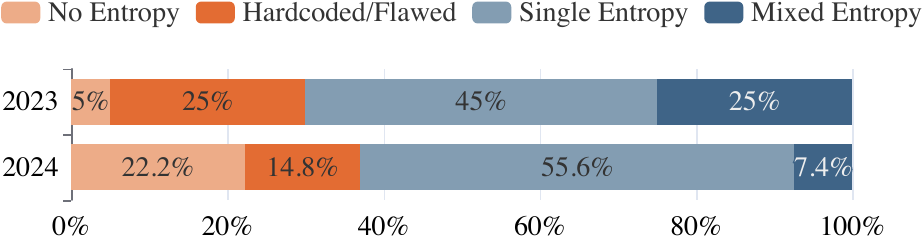}
	\caption{Entropy sources used by teams in the 2023 and 2024.}
	\Description[Entropy sources used by teams in the 2023 and 2024 eCTF.]{Entropy sources used by teams in the 2023 and 2024 eCTF.}
	\label{fig:analysis_entropy}
	\vspace{-5pt}	
\end{figure}

\textbf{\textit{Submission Analysis:}}
We analyzed teams' submissions to understand how they used randomness in their designs, their source(s) of entropy, and shortcomings that might have arisen from their combination of the two.
Figure~\ref{fig:analysis_entropy} shows the entropy source usage by teams in the 2023 and 2024 eCTF.

\noindent\textbf{Teams were not able to maintain unpredictability in their use of randomness.}
In 2023, 1 out of 20 teams neglected the inclusion of entropy in their cryptographic design, while 5 teams relied on hard-coded seeds or entropy generated at build time.
Although some teams updated the seed with each use of the PRNG, the ability of attackers to reset the seed by re-flashing the firmware allowed them to predict the PRNG's output.
In 2024, 18 out of 27 teams utilized the vendor-provided True Random Number Generator (TRNG), while 5 teams did not include randomness in their design.

In addition to hard-coded seeds, some teams opted for entropy sources such as the SysTick counter, built-in timers, or CPU cycle counters. 
While these sources represent an improvement over static seeds, using them without other sources still introduces vulnerabilities. 
Attackers could execute the firmware and repeatedly perform specific operations, cataloging PRNG outputs to construct a comprehensive database. 
A sufficiently extensive database increases the likelihood that the output of a future operation, if conducted at a precise time, could coincide with a database entry, thereby facilitating replay attacks.

\noindent\textbf{Teams utilized different approaches to overcome the challenge and obtain sufficient entropy.}
Addressing the challenge of limited randomness and entropy in microcontroller systems, a viable strategy involves aggregating multiple samples and, when feasible, incorporating various sources of entropy~\cite{grycel2019erhard}. 
This approach mixes random bits across the collected data, thereby maximizing the entropy of randomness.

Several teams adopted this technique by consolidating entropy samples into a pool and subsequently employing a hashing algorithm to derive their seed, mirroring the methodology employed by the Linux PRNG algorithm~\cite{gutterman2006analysis}. 
Teams such as \teamn and \teamd exemplified this practice, while \teamo adopted an alternative approach, hashing values sampled from an internal temperature sensor and combining them through XOR operations.
Both strategies align with NIST recommendations for preserving inputted randomness and mitigating risks associated with insufficient randomness~\cite{NISTentropy}.

Furthermore, diversifying entropy sources can bolster the fault tolerance of the design. 
For instance, \teamn utilized four distinct sources: the drift between the Real Time Clock (RTC) and CPU clock, a device-specific seed, ADC measurements of internal temperature, and readings from uninitialized SRAM. 
Despite the last source being deemed ineffective due to the competition setup, the resulting composite seed remained unpredictable to potential attackers.

\observation{
	Although microcontroller systems usually have limited entropy sources, combining multiple available sources can effectively improve the robustness of the generated randomness.
}

\textbf{\textit{Interviews:}}
We sought to understand the challenges that participants faced with regard to finding and using a reliable source of entropy.
We asked participants about why they chose their entropy source and how they tested its effectiveness.

\noindent\textbf{Participants faced challenges in implementing RNG on the microcontroller system due to limited entropy sources.}
Participant \partd acknowledged low confidence in their RNG implementation, citing insufficient entropy sources and knowledge gaps in microcontroller RNG design:

\begin{quotationsm}
	\textit{``... most implementations of RNG that we could find relied on OS level calls to like urandom ... and we just did not have that on the embedded system.''}
\end{quotationsm}

They struggled with identifying reliable entropy sources, such as SRAM at boot time or temperature sensors, and had difficulties in ensuring adequate entropy was sampled for true randomness:

\begin{quotationsm}
	\textit{``We didn't really know how to ... get solid entropy sources to seed the random number generator to make sure that it's random.''}
\end{quotationsm}

\noindent\textbf{Participants questioned the reliability of the vendor-provided TRNG but did not conduct rigorous tests.}
In 2024, the development board was equipped with a vendor-provided TRNG implementation, which could be accessed through the SDK function calls. 
However, the board's documentation only indicated that the TRNG gathered randomness from various sources on the board, without providing any details about the reliability and quality of the generated randomness~\cite{max78000userguide}.

\parta chose a PRNG seeded by both TRNG and the CPU clock because \textit{``we were not totally sure how much we could trust the TRNG.''} 
Eventually, due to the TRNG's slow performance and the potential for attacks against hardware TRNGs, they decided to mix multiple entropy sources to seed a PRNG.

Participant \partb also didn't trust the TRNG initially and tried to confirm it with the vendor to understand how it was implemented. However, the vendor \textit{``didn't tell us everything,''} and they had to rely on the vendor's assurances.

Most participants did not rigorously test the vendor-provided TRNG; instead, they relied on the vendor's assurances.
According to \partb, since the vendor did not disclose sufficient information and because of the competition schedule, they chose to trust the vendor's certainty and did not perform a robustness test.
Similarly, 18 out of 22 participants admitted to performing only rudimentary tests, such as checking the first few outputs to ensure they appeared random, as mentioned by \partg:

\begin{quotationsm}
	\textit{``We did not test. We just ... printed it ten times and until: hey it's different each time. So we have it good enough.''}
\end{quotationsm}

\noindent\textbf{To further assess the RNG's robustness, several teams conducted ad-hoc tests. However, they struggled to interpret the results, as many tests yielded inconclusive outcomes.}
Participant \partk and \parts's teams experimented with the board under different environmental conditions, such as by freezing the board, \textit{``to see if the TRNG would break or produce predictable values,''} as \parts explained. They also mentioned a basic but effective test they conducted to ensure the TRNG's reliability:

\begin{quotationsm}
	\textit{``If you print out an image of the [output] of the TRNG, it becomes quite obvious ... sometimes you can see boxes or marks where there is some repeatable pattern.''}
\end{quotationsm}

Others, like \partk, applied \textit{``cryptographic distinguishers to see if it's random or not''} and observed no obvious patterns.
\parte's team \textit{``generated around 2 million random numbers from the TRNG''} for preliminary statistical checks.
They all mentioned that their statistical approaches were taken from existing cryptographic research.

\partd stated that they utilized NIST's suite of statistical randomness tests~\cite{rukhin2001statistical} to evaluate the RNG in both years, gathering around a million samples for testing.
They encountered difficulties in interpreting the test results, with many tests returning inconclusive outcomes that did not definitively indicate the RNG's reliability.
In 2024, they also employed the dieharder test suite~\cite{brown2018dieharder}, which is more comprehensive but also more demanding in terms of the sample size required for conclusive results.
\partd admitted:

\begin{quotationsm}
	\textit{``We were more so trusting that the proprietary non-disclosure TRNG was cryptographically secure ... the test was more so just a very quick sanity check to make sure it's not just egregiously bad.''}
\end{quotationsm}

\recommendation{
	\textbf{Developers} should avoid relying on a single source of entropy and rigorously test their RNG implementations.
	\textbf{Vendors} should specify available entropy sources for each board and provide high-level APIs for accessing them.
	Furthermore, if the TRNGs are included, they should disclose the implementation and performance details to ensure transparency.
}

\section{Summary and Future Work}

\textbf{Summary of our findings.}
From our observations, several significant trends related to embedded development and the effectiveness of our security research and curriculum have emerged.

First, secure implementation is an ecosystem-level responsibility: developers ultimately write the code, but their success hinges on educators who provide foundational knowledge, researchers who translate advances into usable techniques, and vendors who ship supportive hardware, SDKs, and libraries.
Although modern languages and recent research offer promising advancements, adopting these technologies still poses conceptual and practical challenges. 
The lack of vendor and library support further hinders the adoption of new technologies, often leaving embedded developers to either create their own solutions or abandon the problem entirely.

Second, MCU developers face a unique set of demands compared to their counterparts working on microprocessor systems. 
They need an in-depth understanding of their specific platform, including how their code interacts with the compiler and underlying hardware. 
This requires not only proficiency in application code but also a thorough knowledge of hardware features and how to use them to enforce security boundaries.
The often opaque and counter-intuitive nature of compiler operations adds to the complexity, placing a substantial burden on developers as they navigate multiple layers of the software/hardware stack.

Lastly, current curricula fall short in preparing students for the specific challenges of embedded development. 
Our study participants were notably puzzled by how to defend against physical threats, highlighting a significant research and practical gap in mitigating embedded-specific threats.

\textbf{Future work.}
Overall, the demands placed on MCU developers are substantial, requiring a unique combination of skills that span software development, computer engineering, and security best practices. 
As the ubiquity, importance, and connectivity of microcontroller systems continue to grow, there is an increasing need for research, education, and tools that can assist developers in navigating these challenges and reducing the likelihood of security vulnerabilities in embedded code.
Future studies could include a systematic review of current educational offerings in secure embedded systems to better contextualize curricular gaps, as well as comparative investigations into whether experienced professionals face similar challenges as less experienced developers.

\section{Related Work}

\textbf{CTF research}. Previous papers on CTF experiences~\cite{vigna2014ten, trickel2017shell, vykopal2020benefits, hannesdottir2021teaching, junior2022learning} have a primary focus on educational purposes.
Vigna et al.~\cite{vigna2014ten} introduced a framework built on a decade's worth of experience in organizing the international Capture the Flag (iCTF), which was further developed to offer a CTF-as-a-service solution~\cite{trickel2017shell}.
Similarly, Vykopal et al.~\cite{vykopal2020benefits} emphasized the advantages of using CTF challenges as hands-on assignments to enhance students' skills.
In addition to the educational benefits, researchers also examined the challenges and obstacles associated with the CTF model itself. 
Such research sheds light on strategies for addressing various challenges and ensuring a successful CTF experience for participants.
Crispin et al.~\cite{cowan2003defcon} described their experience in the Defcon CTF.
Chung et al.~\cite{chung2014learning} discussed methods to overcome the pitfalls and hurdles commonly encountered in organizing CTFs.

Unlike the existing literature, our study uniquely combines submission analysis with participant interviews to provide a dual-perspective understanding of both the technical and human factors influencing security practices.

\noindent \textbf{Secure software development and user studies}.
A series of research efforts~\cite{ruef2016build, parker2020build, votipka2020understanding} has examined the Build it Break it Fix it (BIBIFI)~\cite{BIBIFI} security-oriented programming contest.
They analyzed submissions and interviewed participants to understand the security mistakes made by developers.
Their studies revealed that most vulnerabilities resulted from misunderstandings of security concepts rather than simple mistakes, and factors such as diverse programming experience and code size influence security practices.
However, since the BIBIFI contests focused on microprocessor systems running Linux and may not generalize to more constrained microcontroller contexts, our study fills this gap by examining security challenges specific to microcontroller systems.

In addition to empirical research on developer behavior, there are also studies on the adoption of memory-safe programming languages.
Fulton et al.~\cite{fulton2021benefits} interviewed senior Rust developers to understand the benefits and challenges of adopting Rust in their projects.
They identified drawbacks, including the steep learning curve, limited library support, and concerns about the ability to hire additional Rust developers in the future.
However, their work focused on general software development in industry and did not specifically address embedded systems.
Our study involves early-career developers and offers insights from a different perspective.

Sharma et al.~\cite{sharma2024rust} analyzed over 6,000 embedded Rust projects and surveyed 225 developers to reveal the challenges in embedded Rust development.
They identified several key issues, including limited ecosystem support, inadequate static analysis tools, and difficulties in integrating with C components.
While some of our findings on embedded Rust adaptation align with theirs, our study focuses on the broader challenges faced by developers in microcontroller systems, including both technical and perceptual aspects.

\noindent\textbf{Securing microcontroller systems}.
Many technical solutions have been proposed to protect microcontroller systems against attacks. 
These include privilege separation and compartmentalization~\cite{clements2017protecting, aweke2018usfi, clements2018aces, kim2018securing},
Control-Flow Integrity (CFI) techniques~\cite{zhou2020silhouette, du2022holistic, almakhdhubmurai2020, tan24canary, tan2023sherloc, wang24insect, ma2023dac}, 
randomization methods~\cite{shi2022harm, luo2022faslr},
Return-Oriented Programming (ROP) gadget removal techniques~\cite{kwon2019uxom}, etc.
For a comprehensive review of the research on defensive approaches, please refer to Tan et al.~\cite{tan24sok}.
For real-world firmware analysis, FirmXRay~\cite{wen2020firmxray}, Nino et al.~\cite{nino2024unveiling}, and Tan et al.~\cite{tan24sok} all presented datasets of microcontroller firmware for IoT devices and conducted static analyses to assess the security properties.
Even though the results of these works reveal many concerning issues in real-world firmware, none of them study the perceptual challenges associated with adopting security mitigation for microcontroller systems.

Unlike previous purely technical studies, we examine the primary challenges associated with designing and implementing security mitigation for microcontroller systems from both technical and perceptual perspectives.

\section{Conclusion}

In this study, we investigated the security practices and challenges faced by participants in the 2023 and 2024 MITRE eCTF competitions. 
Through a detailed analysis of competition submissions and interviews with participants, we uncovered both conceptual and practical security gaps in the development of embedded systems. 
Our findings indicate that despite the participants' familiarity with basic security concepts, there is a significant disconnect when applying these mechanisms to embedded systems, compounded by a lack of adequate support for robust, embedded-specific security practices. 
We hope this paper spurs further discussion and collaboration among researchers, vendors, and educators to enhance the state of embedded systems security.

\begin{acks}
The authors extend their gratitude to Ya-Hui Chang for valuable discussions and to Dan Walters for assisting with interview recruitment.
This material is based upon work supported in part by the National Science Foundation (NSF) under grants 2512972, 2508320, 2523436, 2521803, and 2154415, and by an Amazon Research Award, Spring 2025.
Any opinions, findings, conclusions, or recommendations expressed in this material are those of the author(s) and do not reflect the views of the United States Government, any of its agencies, or Amazon.
\end{acks}
\bibliographystyle{ACM-Reference-Format}
\bibliography{acmart}


\end{document}